# A LOW COMPLEXITY SPACE-FREQUENCY MULTIUSER SCHEDULING ALGORITHM

*Ana I. Pérez-Neira[1,2], Pol Henarejos[1], Velio Tralli[3], Miguel A. Lagunas[1,2]*

email: anuska@gps.tsc.upc.edu, pol@redyc.com, vtralli@ing.unife.it, m.a.lagunas@cttc.es
(1) Dept. of Signal Theory and Communications - Universitat Politècnica de Catalunya (UPC) – Spain
(2) Centre Tecnològic de Telecomunicacions de Catalunya (CTTC) - Spain
(3) ENDIF - Engineering Department University of Ferrara - CNIT - Italy

**ABSTRACT**

This work presents a resource allocation algorithm in K-user, M-subcarrier and NT-antenna systems for on-line scheduling. To exploit temporal diversity and to reduce complexity, the ergodic sum rate is maximized instead of the instantaneous one. Dual optimization is applied to further diminish complexity together with a stochastic approximation, which is more suitable for online algorithms. Weighted sum rate is considered so that users can be either prioritized by higher layers or differentiated by proportional rate constraints. The performance and complexity of this algorithm is compared with well-known benchmarks and also evaluated under real system conditions for the MIMO Broadcast channel.

## 1. INTRODUCTION

In a Multi-User MIMO (MU-MIMO) spatial multiplexing scheme, multiple users are scheduled in the same resource block. MU-MIMO is a promising way to increase system throughput and there is a growing interest on the topic as [1,2,3, 4] shows. Recently attention has been paid to the combination of spatial diversity multiple access systems and frequency domain packet scheduling [5,6,7,8,9]. Specifically, in [6] the authors present a low complexity sum-power constraint iterative waterfilling that is capacity achieving. It improves the convergence of [3] and is probably convergent. In [8] the authors address the problem of feedback reduction. The present paper aims at both, low complexity and reduced feedback. In contrast to [7], in order to further reduce complexity for on line implementation we follow a dual decomposition strategy and a stochastic approximation. In order to reduce feedback load the paper resorts to opportunistic strategies that solve the spatial scheduling. More specifically, an efficient algorithm for optimal beam subset and user selection is performed to find the best trade-off between the multiplexing gain and the multiuser interference in the opportunistic scheme when the number of users is not high. In summary, this paper proposes a joint spatial and frequency scheduler that allows on-line implementation and only requires partial or low feedback and a low-complexity implementation.

This paper is organized as follows. The space-frequency scheduler is formulated in Section 2 and the distributed scheme that is proposed based on dual optimization is presented in Section 3. Section 4 explains the low complexity ergodic algorithm, together with an evaluation of its complexity. The numerical results are presented in Section 5, and Section 6 concludes the paper.

## 2. PROBLEM FORMULATION

We consider an OFDMA scenario with M subcarriers and K users. Each user $k$ is single antenna and receives simultaneously $N_T$ signals, which can come from different spatial locations, antennas or beams. Only one of the $N_T$ signals is intended for user $k$. The received signal by user $k$ on the subcarrier $m$ is given by

$$y_{k,m} = a_{k,m}\sqrt{p_{k,m}}\alpha_{k,m,k_q}s_{k,m} + \sum_{s \neq k} a_{s,m}\sqrt{p_{s,m}}\alpha_{k,m,s_q}s_{s,m} + w_{k,m} \quad (1)$$

where, $\mathbf{a} = \{a_{k,m}\}$ is the set of binary allocation variables, i.e. $a_{k,m} = 1$ if user $k$ is scheduled on frequency $m$, $a_{k,m} = 0$ otherwise, and $\mathbf{p} = \{p_{k,m}\}$ is the set of allocated powers, $s_{k,m}$ is the information signal of user $k$ through frequency $m$, $E\{|s_{k,m}|^2\} = 1$. Finally, $\alpha_{k,m,k_q}$ denotes the equivalent channel seen by the $k$th user at frequency $m$ with respect to the $q$th beam, antenna or transmitter associated with user $k$. For instance, in the case of a MISO (Multiple input single output) broadcast channel, $|\alpha_{k,m,k_q}|^2 = c_{k,m,k_q} = |\mathbf{h}_{k,m}^T \mathbf{b}_{m,k_q}|^2$, where $\mathbf{b}_{m,k_q}$ is the beamforming vector that is associated with user $k$ and that is obtained from the set of beams $k_q \in \{1,...,N_T\}$. In this case the number of interference terms in (1) is equal to $N_T-1$. From a viewpoint of information theory, the model in (1) could correspond either to a broadcast channel or to an interference channel with $N_T$ transmitters and K receivers. In spite of the big gains in spectral efficiency that can be obtained by incorporating multiantenna transmission to a multicarrier system, an evident drawback of this scenario is the increased design complexity. In other words, multiantenna, multiuser and multicarrier channels significantly increase the set of design parameters and degrees of freedom at the PHY layer. In this work, the focus is on the optimization of the PHY layer parameters with low complexity burden. Concerning the optimality criteria, we consider the problem of rate maximization in (2) with power constraints and also proportional rate constraints.

$$\max_{\mathbf{a},\mathbf{p}} R(\mathbf{a},\mathbf{p}) = E_\gamma \left\{ \sum_{k=1}^{K} \sum_{m=1}^{M} \log_2 \left(1 + \gamma_{k,m,k_q}(\mathbf{a},\mathbf{p})\right) \right\}$$

$$s.t. \quad E\left\{ \sum_{m=1}^{M} \log_2 \left(1 + \gamma_{k,m,k_q}(\mathbf{a},\mathbf{p})\right) \right\} \geq \phi_k R, \quad k=1,...,K$$

$$\sum_{k=1}^{K} \phi_k = 1 \quad (2)$$

$$E_\gamma \left\{ \sum_{m=1}^{M} \sum_{k=1}^{K} a_{k,m} p_{k,m} \right\} \leq \overline{P}$$

$$a_{k,m} \in \{0,1\} \quad m=1,...,M$$

$$p_{k,m} \geq 0 \quad m=1,...,M \quad k=1,...,K$$

with

$$\gamma_{k,m,k_q}(\mathbf{a},\mathbf{p}) = \frac{a_{k,m} p_{k,m} c_{k,m,k_q}}{\sigma^2 + \sum_{s \neq q}^{N_T} a_{s,m} p_{s,m} c_{k,m,k_s}} \quad (3)$$

where **a** and **p** are vectors whose components are $a_{k,m}$ and $p_{k,m}$, respectively. $\gamma_{k,m,k_q}(\mathbf{a},\mathbf{p})$ is the SINR (Signal to Interference and Noise Ratio) of user $k$ at frequency $m$ and associated with beam $q$ and $c_{k,m,k_q}$ denotes the equivalent channel power gain seen by the $k$th user at frequency $m$ with respect to the $q$th beam, antenna or transmitter. The formulation of the sum rate in (2) indicates that at each frequency up to $N_T$ transmissions can be spatially multiplexed. We assume that $\gamma_{k,m,k_q}(\mathbf{a},\mathbf{p})$ are known by the $N_T$ transmitters by means of partial channel feedback. For instance, this would be the case of a broadcast channel where the Base Station (BS) has perfect SINR feedback. Other possible scenario is that of $N_T$ BS's in a cellular system; in this case the assumption would be that all BS know the equivalent channel magnitude. $\phi_k$ are the weights that allow prioritizing the users. The problem to solve deals with scheduling of users and powers, the spatial precoder is fixed and part of the initial conditions of the problem.

Rate optimization is a reasonable choice for utility, reflecting the various coding rates implemented in the system. We assume an idealized link adaptation protocol. Proportional rate constraints allow a more definitive priorization among the users, which is quite useful for service class differentiation. Theoretically, this formulation also traces out the boundary of the capacity region similar to the weighted sum-rate maximization. The main difference is that it actually identifies the points on the capacity region boundary that satisfy the rate proportionally constraints. Furthermore, the max-min rate formulation is a special case of this formulation, i.e., when $\phi_1 = ... = \phi_K$. Finally, by enforcing the average power constraint we allow instantaneous power levels to exceed the average power when necessary. Sum power constraint is needed in scenarios such as BC channel, but it is not usually imposed in multi-cell scenarios.

Note that ergodic optimization is considered because of twofold: i) it reduces the complexity of the resulting algorithm and ii) it incorporates the time dimension in the resulting resource allocation. In other words, in the case of instantaneous rate allocation only, the OFDMA algorithms are re-run every symbol (or several symbols). In this paper, we can capture the idea of "time slot allocation" by using the ergodicity assumption, and determine power allocation functions that are parameterized by the channel knowledge.

Note also that if there is no frequency structured components or noise, the maximal sum rate signaling does not require introducing correlation between subcarriers (cooperative subcarrier transmission or joint frequency-space processing). Therefore, the problem is separable across the subcarriers, and is tied together only by the power constraint. In these problems, it is useful to approach the problem using duality principles. In addition, the utility function is non convex and by solving the dual problem and formulating it as a canonical distributed algorithm [10], the algorithm is simplified and also convergence to the globally optimal rate allocation can be achieved.

Finally and as notational convention vectors are set in boldface.

### 3. DUAL OPTIMIZATION

The proposed algorithm is based on a dual optimization framework. In other words, it is based on a Lagrangian relaxation of the power constraints and (possibly) rate constraints. This relaxation retains the subcarrier assignment exclusivity constraints, but "dualizes" the power/rate constraints and incorporates them into the objective function, thereby allowing us to solve the dual problem instead. This dual optimization is much less complex as we explain next.

To derive the dual problem we first write the Lagrangian. In order to simplify notation we define

$$r_k \triangleq \sum_{m=1}^{M} E_\gamma \left\{ \log_2 \left(1 + \gamma_{k,m,k_q}\right) \right\}$$
$$\hat{p}_k \triangleq \sum_{m=1}^{M} E_\gamma \left\{ a_{k,m} p_{k,m} \right\} \quad (4)$$

where we do not explicitly write the dependence of $r_k$ and $\hat{p}_k$ on the optimization variables **a**, **p**. Based on this definitions, the Lagrangian is

$$L = R(1 - \boldsymbol{\mu}^T \boldsymbol{\phi}) + \lambda \overline{P} - \lambda E_\gamma \left\{ \sum_{k=1}^{K} \hat{p}_k \right\} + \sum_{k=1}^{K} \mu_k r_k \quad (5)$$

$\lambda$, $\boldsymbol{\mu}$ are the dual variables (also called prices) that relax the cost function, $R = \sum_{k=1}^{K} r_k$ and $\overline{P}$ is the power constraint.

Focusing on the first term in the maximization, we observe that if $(1 - \boldsymbol{\mu}^T \boldsymbol{\phi}) > 0$ then the optimal solution would be $R^* = \infty$, since R is a free variable. This is clearly an infeasible solution for ergodic sum rate. Furhtermore, if $(1 - \boldsymbol{\mu}^T \boldsymbol{\phi}) < 0$ then the optimal solution would be $R^{opt} = 0$. Thus, we would like to constrain the multiplier to satisfy $\boldsymbol{\mu}^T \boldsymbol{\phi} = 1$. Thus, (5) can be simplified to

$$L = \lambda \overline{P} - \lambda E_\gamma \left\{ \sum_{k=1}^{K} \hat{p}_k \right\} + \sum_{k=1}^{K} \mu_k r_k \quad (6)$$

Note that the weights $\mu_k$ are the dual multipliers that enforce the proposed rate constraints. Additivity of the utility and linearity of the constraints lead to the following Lagrangian dual decomposition into individual user terms

$$L = \sum_k L_k + \lambda \overline{P} \qquad (7)$$

where, for each user k,

$$L_k = \sum_k \mu_k r_k - \lambda_k \sum_k \hat{p}_k \qquad \lambda_k = \lambda \qquad (8)$$

only depends on local rate $r_k$ and the prices $\lambda$, $\mu$.
The dual function $g(\lambda, \mu)$ is defined as:

$$g(\lambda, \mu) = \max_{\mathbf{a},\mathbf{p}} \left( \sum_k L_k(\mathbf{a},\mathbf{p},\lambda_k,\mu) + \lambda \overline{P} \right) = \sum_k L_k^*(\mathbf{a}^*,\mathbf{p}^*,\lambda_k,\mu) + \lambda \overline{P} \qquad (9)$$

Evidently, this dual problem involves only K+1 variables and it is, therefore much easier to solve than the primal problem. Moreover, the maximization in (9) can be conducted parallel by each user, as long as the aggregate link price λ is feedback to source user *k*. Note that if if there were no global constraint, as it is the case in multicell systems where $p_{k,m}$ stands for the power that Base Sation *k* has to allocate in frequency *m* and there is only per BS power constraints, the problem is further simplified.

The dual problem is defined as:

$$\begin{aligned} &\min g(\lambda, \mu) \\ &s.t. \quad \lambda \geq 0, \mu \in D \quad D = \{\mu \geq 0, \mu^T \phi = 1\}. \end{aligned} \qquad (10)$$

Since $g(\lambda, \mu)$ is the pointwise supremum of a family of affine functions in $\lambda$, $\mu$, it is convex and (10) is a convex minimization problem (even if the primal is not a concave maximization problem).

Since $g(\lambda)$ may be non differentiable, an iterative subgradient method can be used to update the dual variable λ to solve the dual problem. The computation of the subgradient requires knowing the individual weighted ergodic rates per user. Note that the "weights" in this case are no longer predetermined constants, but are effectively the multipliers that enforce the proportional rate constraints. From an initial guess $\lambda^o$ and $\mu^o$, the subgradient method generates a sequence of dual feasible parts according to the iteration

$$\lambda^{i+1} = \left[ \lambda^i - s^i g_\lambda^i \right]^+ \qquad \mu^{i+1} = \Pi_D \left[ \mu^i - s^i g_\mu^i \right] \qquad (11)$$

Where $g_\lambda^i$ denotes the subgradient of $g(\lambda^*(\mu^i), \mu^i)$ with respect to λ

$$g_\lambda^i = \overline{P} - E_\gamma \left\{ \sum_k \hat{p}_k^* \right\} \qquad (12)$$

and $s^i$ is a positive scalar step-size. $g_\mu^i$ denotes the subgradient of $g(\lambda^*(\mu^i), \mu^i)$ with respect to μ

$$g_\mu^i = \overline{\mathbf{R}}^i - \phi \overline{R}^i$$

with $\left[ \overline{\mathbf{R}}^i \right]_k = \sum_m E_\gamma \left\{ \log_2 \left( 1 + \gamma_{k,m,k_q}(\mathbf{a}^*,\mathbf{p}^*,\lambda^i,\mu_k^i) \right) \right\} \qquad (13)$

$$\overline{R}^i = \sum_k \left[ \overline{\mathbf{R}}^i \right]_k$$

Finally, $\Pi_D[.]$ denotes projection onto the set D.

Concerning convergence, for a primal problem that is a convex optimization, the convergence is towards a global optimum. Otherwise, global maximum of non concave functions is an intrinsically difficult problem on non convex optimization. In [10] the authors show that the sequence of the maximization of (8) and the computation of (11) forms a canonical distributed algorithm that solves (2) and the dual problem (10). Even for non concave utilities the canonical distributed algorithm may still converge to a globally optimal solution if $L_k^*$ is continuous at optimum $\lambda^*$. Based on this property, an analytical proof of convergence for the algorithm that is proposed next is left for further work. Simulation results have proved good convergence for it.

## 4. ALGORITHM

In the rest of the paper we deal with the specific case of Broadcast (BC) channel, where the global power constraint is needed. The maximization of (9) could have been formulated only with respect to the powers, **p**. In this way, whenever any of the optimal components $p_{k,m}^* = 0$, this would mean that user *k* should not be scheduled in frequency *k*. In addition to the complexity of this multiuser frequency power allocation problem, note that $p_{k,m}$ depends on the spatial channel at frequency *m*, which, in the case of the BC channel, depends on the spatial precoder that is associated with user *k*. The purpose of this work is to design a low complexity scheduler; this fact motivates the simplification of the complex space-frequency multiuser scheduler that has been described by using the multibeam opportunistic scheme [11]. In this case, the Base Station uses a set of orthonormal beams that are associated with the users depending on their reported SINR. This scheduler is designed such that it works without interacting with power allocation and dual optimization. The problem formulation of (9) accounts for this explicit user scheduling by incorporating the discrete variables *a*. As shown in Fig. 1, the first step in the proposed algorithm is the spatial scheduler, which obtains *a* in a low complexity way, as it is described in Section A.

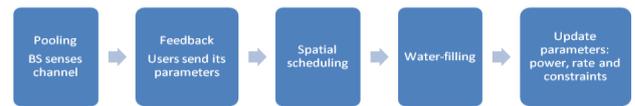

**Fig. 1**. Block diagram of the proposed algorithm.

Once user *k* has been selected by being associated with beam *q* at each frequency *m*, the next step in the algorithm (see Fig. 1) is the power allocation, which should be derived from the equation

$$\frac{\partial L_k}{\partial p_{k,m}} = 0 \qquad \forall k \qquad (14)$$

Assuming the interference with constant power $\overline{P}$, a suboptimal solution is provided by the waterfilling in (15)

$$p_{k,m}^* = \left[ \frac{\mu_k}{\lambda \ln(2)} - \frac{1}{\gamma_{k,m,k_q}^1} \right]^+ \qquad (15)$$

where

$$\gamma^1_{k,m,k_q} = \frac{c_{k,m,k_q}}{\sigma^2 + \sum_{s \neq q}^{N_T} a_{s,m} P c_{k,m,k_s}} \quad (16)$$

The dual variables are obtained by solving the dual problem (10) with low complexity in a stochastic way. The detailed waterfilling procedure is described in Section C.

*A. Spatial scheduler*

The spatial scheduler is obtained for each frequency $m$ by introducing a simplified SINR that considers uniform power allocation (i.e. no interaction neither with the primal variables **p** nor with the dual variables $\lambda$, $\mu$). The proposed spatial scheduler is based on the multibeam opportunistic strategy, which considers an orthonormal random beamforming set as precoder and assigns users to beams based only on SINR feedback. In order to counteract the losses that opportunistic schemes present when the number of users is moderate or low we extend the beam and user set optimization proposed in [12] to OFDMA. Let $Q_m \in \{1,...,N_T\}$ denotes the number of users served or active beams at $m$th frequency bin, $U_m^{(Q_m)} \subset \{1,...,K\}$ and $S_m^{(Q_m)} \subseteq \{1,...,N_T\}$ are the user and beam set, respectively, with $Q_m$ elements without repetition. For the BC the SINR in (3) is equivalent to

$$\gamma^{(Q_m)}_{k,m,q} = \frac{p_{k,m} |\mathbf{h}^T_{k,m} \mathbf{b}_{m,q}|^2}{\sigma_n^2 + \sum_{\substack{q' \neq q \\ q' \in S_m^{(Q_m)}}} p_{q',m} |\mathbf{h}^T_{k,m} \mathbf{b}_{m,q'}|^2} \quad m \in S_m^{(Q_m)} \quad k \in U_m^{(Q_m)} \quad (17)$$

where $\mathbf{b}_{m,q}$ is $q$th beam at $m$th frequency and $\sigma_n^2 = \sigma_w^2/P$ is the noise variance. In (17) there are $\binom{K}{Q^m} Q^m!$ permutations of $\gamma^{(Q_m)}_{k,m}$ for $S^{(Q_m)}$. The optimal beam subset $S_m^*$ and user subset $U_m^*$ at $m$th frequency bin are obtained by (9) using exhaustive search. However we apply a suboptimal approach to reduce the complexity. Next, we assume that equal power P is allocated among beams and subcarriers, $j = 1,...,\binom{N_T}{Q_m}$ is the index of all beam combinations.

Fixing $j$th combination, the optimal user satisfies

$$k^*_{m,j,q} = \arg\max_k \gamma^{(Q_m)}_{k,m,j,q} \quad (18)$$

where subindex $k_q$ in (3) has become indexes $k,j,q$ in (18) due to the beam and user search. The optimal value of (18) is added to $U^{(Q_m)}_{j,m}$. For each frequency $m$, optimal $j$th index is found by expression

$$j^* = \arg\max_{1 \leq j \leq \binom{N_T}{Q_m}} \sum_{q=1}^{Q_m} \log_2\left(1 + \gamma^{(Q_m)}_{k^*_{m,j,q},m,j,q}\right) \quad (19)$$

Hence, the optimal user and beam set that serves $Q_m$ users simultaneously are $U_m^{*(Q_m)} = U^{(Q_m)}_{j^*,m}$ and $S_m^{*(Q_m)} = S^{(Q_m)}_{j^*,m}$. Finally, optimal $Q_m$ is given by

$$Q_m^* = \arg\max_{1 \leq Q_m \leq N_T} \sum_{k \in U_m^{*(Q_m)}} \sum_{q \in S_m^{*(Q_m)}} \log_2\left(1 + \gamma^{(Q_m)}_{k,m,j^*,q}\right) \quad (20)$$

Once this spatial scheduling is finished for each frequency $m$, we simplify indexes $\binom{Q_m^*}{k^*_{m,j,q},m,j^*,q} \triangleq k,m,k_q$ as they are used in (3), ready to be applied in the frequency waterfilling and dual optimization step. The corresponding user selection variable $a_{k,m}$ is set to 1 if the user has been chosen.

Regarding fairness, this solution has the drawback of scheduling users on the available beams and frequencies by only looking at channel gains. In this way, users with good channel conditions, i.e. users located near the base station with a small path loss, tend to monopolize channel resources. The lack of resources for weak users may constrain the behavior of dual optimization. An improved spatial scheduler can be designed by considering the maximization of

$$\max_{k_q} \sum_k \sum_q \mu_k \log_2\left(1 + \gamma_{k,m,k_q}\right) \quad \forall m \quad (21)$$

by using the equal power approximation of $p_{k,m}$, i.e. $p_{k,m}=P$. This can be simply implemented by inserting $\mu_k$ in (19) and (20). This scheduler interacts with the dual optimization algorithm. It releases beams and frequencies to users according to rate constraints, but preserves both the light requirements on feedback parameters and the distributed implementation.

Finally, instead of MOB other spatial precoders (such as Zero Forzing) can be used, however, at the expense of complexity increase in the feedback.

*B. Feedback complexity*

The number of feedback parameters to be transmitted to the BS by each $k$th user are $N_T$x$M$ and they correspond to all possible values of $|\mathbf{h}^T_{k,m}\mathbf{b}_{m,q}|^2$ for $q=1,...,N_T$ and $m=1,...,M$. This amount can be reduced by fixing $Q_m$ previously. Users must know the value $Q_m$. In this case, feedback is reduced to 3x$M$ and they correspond to $q^*$, $j^*$, best beam and best permutation, and $\gamma_{k,m,j^*,q^*}$. That is because fixing $Q_m$ causes the number of permutations is known a priori by all users and it is fixed. Thus, each user can compute $\gamma_{k,m,j^*,q^*}$ and send it to BS, jointly with indexes. In addition, fixing $Q_m = Q = N_T$ the amount of feedback is also reduced to 2x$M$, $q^*$ and $\gamma_{k,m,1,q^*}$ since $j=1$.

Finally, depending on the delay spread of the channel, the number of parameters to feedback can be further reduced by frequency grouping or chunk processing [5].

*C. Frequency power allocation*

Once the user selection problem has been solved, **a** is known in (2), and the next step is to obtain the solution for the power allocation **p** in (12).

To compute (12), ergodic maximization through stochastic approximation is introduced [14]. In other words, maximization occurs through time or iterations. At $n$th iteration, power assigned to $k$th user at $m$th carrier becomes

$$p^*_{k,m}[n] = \left[\frac{\mu_k[n]}{\lambda[n]\ln(2)} - \frac{1}{\gamma^1_{k,m,k_q}[n]}\right]^+ \quad (22)$$

where $[x]^+ = \max(x,0)$.

Per-user rate and total power are given by

$$R_k[n] = \sum_{m \in M} R_{k,m}\left(\gamma_{k,m,k_q}(p^*_{k,m}[n])\right)$$
$$P[n] = \sum_{m \in M}\sum_{k \in K} p^*_{k,m}[n] \quad (23)$$

Finally, $\lambda$, $\mu$ are updated using subgradient method as in (11)-(13) but with a stochastic approximation. In fact, these parameters are given by expressions

$$\lambda[n+1] = \left[\lambda[n] - \delta(\bar{P} - P[n])\right]^+$$
$$\mu[n+1] = \Pi_\mathcal{D}\left(\mu[n] - \delta(\mathbf{R}[n] - \phi R[n])\right) \quad (24)$$

where $\Pi_\mathcal{D}$ is the projection onto set $\mathcal{D} = \{\mu \geq \mathbf{0} | \mu^T \phi = 1\}$, $\mathbf{R}[n] = [R_1[n] \cdots R_K[n]]^T$ and $R[n] = \sum_{k \in K} R_k[n]$. To obtain a good performance, a suitable values could be $\lambda[0] = 1$, $\mu[0] = \dfrac{\phi}{\phi^T\phi}$ and $\delta = 0.01$.

Subgradient search methods have been used to obtain the solution for the dual variables. As it is important to be able to perform resource allocation in real-time, we obtain an on-line adaptive algorithm by performing the iterations of the subgradient across time.

### D. Complexity

Using ergodic sumrate relaxes complexity since constraints are not instantaneous but ergodic. Note that though objective is ergodic, feedback parameters contain instantaneous information. This algorithm has several stages. First of all there is the pooling stage. During this step, BS sends a pilot signal through all beams and only one beam is active. The complexity in the beamforming is $O(MN_T)$.

Next step is computing all $\gamma$ parameters. Its complexity depends on fixing $Q_m$. Leaving it as a free parameter, complexity is $O(MK2^{N_T}N_T^2)$. Otherwise, fixing it, complexity becomes $O(MK2^{N_T}N_T)$. Moreover, adjusting $Q_m = Q = N_T$ it is reduced to $O(MKN_T)$.

Finally, there is the power allocation stage, water-filling has complexity $O(MK)$ and it is followed by $O(K)$ updates for the rates, power and multipliers.

In general, complexity could be very low, as $O(MKN_T)$, or higher, as $O(MK2^{N_T}N_T^2)$, depending on how optimum is desirable.

Next table shows algorithm step-by-step and its complexity. Note that in step 2 there are three possibilities: 2.a has $N_T \times M$ parameters of feedback; 2.b, 3xM and 2.c, 2xM.

Other solutions such as [13] find the optimal bound of capacity rate at cost of complexity, $O(K^2 M \log N)$. Others have less complexity, as $O(M)$ in [8], but they lose in performance. See fig. 4 for more details in the comparison.

## 5. RESULTS

We organize numerical results in two parts: the first part illustrated in section 1 refers to a simple cellular scenario and has the aim of showing the main behavior of the algorithm; the second part refers to a more realistic scenario.

| Step | Complexity |
|---|---|
| 1. Pooling: BS transmits pilot signal to sense each equivalent channel. | $O(MN_T)$ |
| 2.a Non-fixed $Q_m$ <br> Feedback: each user sends $\left|\mathbf{h}^T_{k,m}\mathbf{u}_{m,q}\right|^2$ <br> BS schedules users spatially | $O(MN_T)$ <br><br> $O(MK2^{N_T}N_T^2)$ |
| 2.b Fixed $Q_m$ <br> Feedback: each user sends $q^*, j^*, \gamma_{k,m,j^*,q^*}$ <br> BS schedules users spatially | $O(M)$ <br><br> $O(MK2^{N_T}N_T)$ |
| 2.c $Q_m = N_T$ <br> Feedback: each user sends $q^*, \gamma_{k,m,1,q^*}$ <br> BS schedules users spatially | $O(M)$ <br><br> $O(MKN_T)$ |
| 3. Water-filling | $O(MK)$ |
| 4. Updating parameters $\lambda, \mu$ | $O(K)$ |

**Table 1.** Algorithm step-by-step and its complexity.

### A. Results for simple scenario

All simulations in this scenario consider M=64 subcarriers, power constraint $\bar{P}=10dB$ and power parameter P=1 in equation (16). The channel model includes normalized Rayleigh fading and does not take care of path-loss or shadowing components. All carriers have frequency spacing of 1Hz. All users are located at same distance from BS. A linear array of NT antenna is considered at the base station and distance between sensors is $0.5\lambda$.

Fig. 2 shows how every user converges to its weight with few iterations. Note that user rates are normalized with sumrate. Fig. 3 shows power convergence to average power constraint. Hence the good convergence properties of dual optimization algorithm are confirmed by the results.

In order to compare the spatial scheduling algorithm with other algorithms of the literature, we show in Fig. 4 the rate region for two users, compared with those obtained with the algorithms in [13] (DPC) and [8] (KOU), and to uniform power allocation strategy (UPA). Note that [13] has higher complexity and fixes a theoretical maximum sum-rate that can be achieved, whereas [8] has lower complexity, but worse performance.

Fig. 5 shows the computation complexity in terms of simulation time for different system configurations and strategies of choosing $Q_m$. Note how complexity increases with the usage of dynamic $Q_m$. In order to compare the complexity with that of [13], based on DPC, we provide the following table that shows the computation complexity for K=8 and K=32 users, and $N_T$=2. Time is expressed in seconds. This gives a clear idea of the complexity of [13], $O(K^2 M \log N)$, and how it increases with number of users.

|  | K=8 | K=32 |
|---|---|---|
| [13] | 283 | 3349 |
| Our algorithm <br> (Dyn $Q_m$ / $Q_m = N_T$) | 28 / 31 | 50 / 62 |

**Table 2.** Complexity comparison with [13].

When compared to solution [8], for $N_T=2$ and $K=2$, our algorithm requires roughly the same computation time, but achieves best performance, as shown in Fig. 4.

Finally, Fig. 6 shows the impact on sum-rate of using a dynamic $Q_m$ or a fixed $Q_m$. Dynamic choice of $Q_m$ is useful for a large number of antennas, whereas the other choice is good for few users.

*B. Results in a realistic scenario*

In this section we present and discuss simulation results obtained for a scenario which incorporate some characteristic aspects of practical application in next generation wireless systems, In fact, 3GPP-LTE and WiMAX employ OFDMA as their main multiple access mechanism (although other options are also defined in the standards).

We are considering here a single cell of the downlink of an OFDM wireless system with M=128 subcarriers working on a bandwidth of 1.25 Mhz. Base station is equipped with multiple antennas. The system is TDD and it is assumed that 2/5 of frame interval is used for downlink transmission. The CSI coming from users is updated every 10ms. Two options for user distribution are considered: in the first option the users have a position which is uniformly distributed in circular area of radius 500m; in the second option the users are placed at the same distance of 250 m from the base station.

Channel model includes path loss, correlated shadowing (not present in the second option for user distributions) and time and frequency correlated fast fading. Path loss is modeled as a function of distance as L(db)= k0 + k1 log(d) (k1=40, k1=15.2 for results). Shadowing is superimposed to path-loss, with classical lognormal model (sigma= 6 dB) and exponential correlation in space (correlation distance equal to 20m). Fast fading on each link of the MIMO broadcast channel is complex Gaussian, independent across antennas and is modeled according to a 3GPP Pedestrian model [11]. This model has a finite number of complex multipath components with fixed delay (delay spread around 2-3 microseconds) and power (average normalized to 1). Time correlation is obtained according to a Jakes' model [12] with given Doppler bandwidth (6 Hz in the results). At the base station orthogonal beamforming is adopted, where beam vectors change randomly at each frame. In the simulated system the total average power constraint is fixed to 1W.

Note that, in realistic conditions, channel variations in time due to Doppler effects have a non negligible impact on the feedback quality. In fact, at the scheduling time n the algorithm uses feedback parameters measured at time n-1, which can be changed in the meanwhile. Therefore, due to outdated feedback the transmission at the scheduled rate may fail sometimes. This aspect is left for future investigation

In the first two figures, fig. 7 and fig. 8, the dynamic behavior of sum-rate and total allocated power is illustrated for a system with 10 users in fixed position at distance 250 m. In the scheduling algorithm equal weights $\phi_k = 1/K$ are used. We note that although total power and sum rate change frame by frame due to fast channel variations, the algorithm for dual variable optimization correctly tracks the constraint on the average power. We also observed from a wide set of results that the range of variations enlarges when the users have different path-loss, but again the algorithm tracks correctly the average.

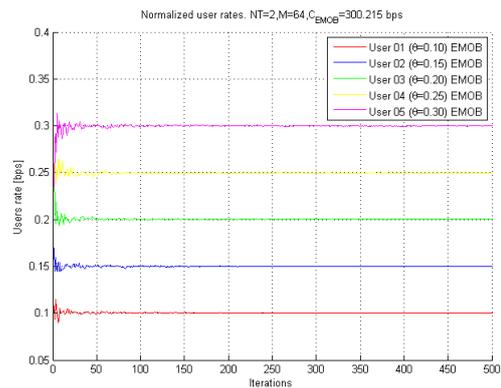

**Fig. 2**. Different rates for 5 users using this algorithm

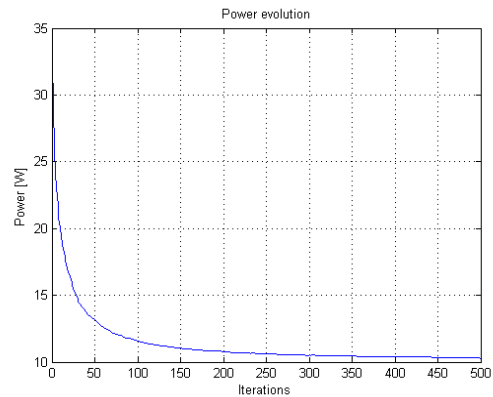

**Fig. 3**. Power evolution vs. iterations.

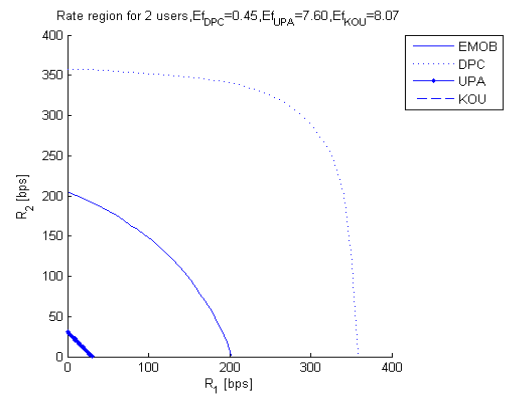

**Fig. 4**. Rate region for two users. DPC from [8], and KOU from [10] are plotted jointly with this algorithm EMOB. UPA is also plotted. Note that UPA and KOU are practically coincident.

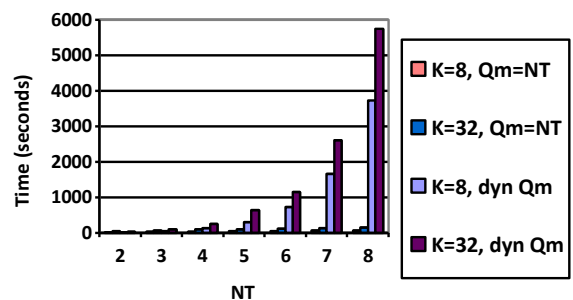

**Fig. 5**. Simulation time in seconds for K=8 and K=32, dynamic $Q_m$ or fixed $Q_m$.

In spite of rate and power variations, we also checked the robustness of algorithm to ensure fair average rate allocation among users. This is shown in the following two histograms in fig. 9 and fig. 10, which illustrates the distribution of average user rates in a system with 10 users, 3 antennas and 3 different user classes (class 1: weight 0.5/K - class 2: weight 1/K - class 3: weight 1.5/K). Fig. 9 refers to equal distance users at 250 m, whereas fig. 10 refers to uniformly distributed users in a circular area of radius 500 m. We can note that the algorithms are quite fair to assign rates to different users, even when they have with different weights and different path-loss conditions.

Finally, fig. 11 shows sum rate and user rate vs. number of user, in a system with 3 antennas, users at equal distance 250 m from BS, and 3 different user classes (class 1: weight 0.5/K - class 2: weight 1/K - class 3: weight 1.5/K). We note that sum rate increases with the number of users, meaning that the scheduling algorithm capture the available multiuser diversity while preserving average rate fairness. Per-user rate decreases since the sum-rate needs to be shared among an increasing number of users.

## 6. CONCLUSIONS

This paper has presented a low complexity space-frequency scheduler that allocates power among users. Ergodic objective and ergodic constraints are purposed to relax complexity. Moreover many strategies had been presented and low complexity has been explained. In addition, weights are purposed in order to set rate priorities or several QoS. Finally, some benchmarks are presented to compare the performance. Aspects such as robustness to imperfect CSIT, discrete rate allocation, modification of the algorithm to incorporate jointly encoded sub-channels (e.g. space-time codes) and cross-layer design for user scheduling improvement are possible topics for further research. The space-frequency multiuser scheduler has been presented in a general formulation such that the proposed distributed strategy (as a result of the dual optimization and opportunistic user selection) together with the low complexity of the proposed ergodic scheduler can be applied to different space-frequency scheduling scenarios.

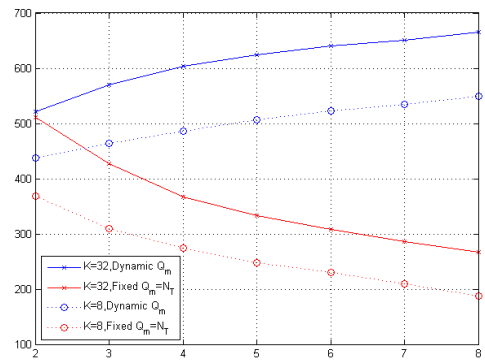

**Fig. 6**. Contribution to sum-rate of dynamic $Q_m$ or fixed $Q_m$. plotted for K=8 and K=32 users with different number of antennas.

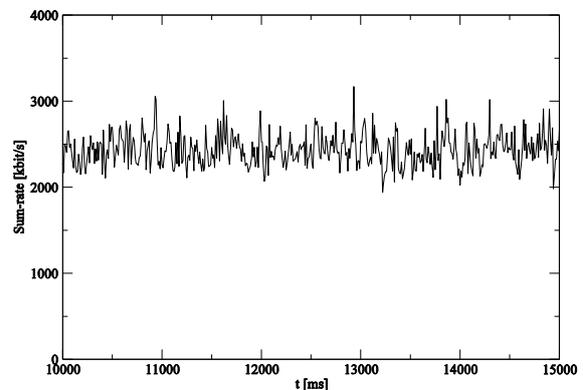

**Fig. 7**. Dynamic behavior of sum-rate in a cell with 10 users in fixed positions at distance 250m from BS.

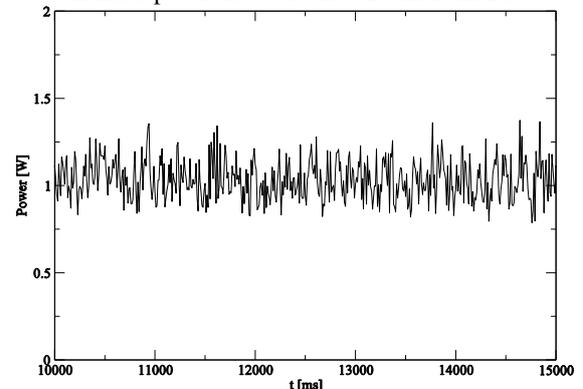

**Fig. 8**. Dynamic behavior of total allocated power in a cell with 10 users in fixed positions at distance 250m from BS.

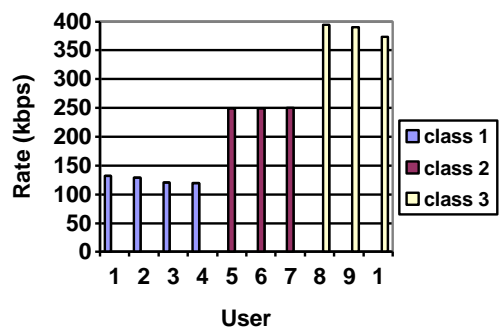

**Fig. 9**. Per-user average rate distribution in a system with 10 users at equal distance 250 m from BS, belonging to 3 different user classes.

## 8. ACKNOWLEDGMENTS

This work was supported by the European Commission under project NEWCOM++ (216715), Optimix (Grant Agreement 214625) and by Spanish Government TEC2008-06327-C03-01. The work has been done during the 6 months stay of A. Perez-Neira at ACCESS/Signal Processing Lab, KTH (Stockholm).

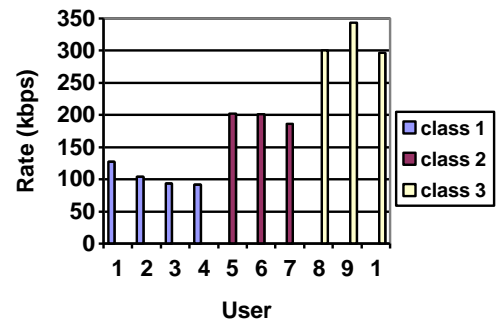

**Fig. 10**. Per-user average rate distribution in a system with 10 users uniformly distributed in a cell of radius 500 m, belonging to 3 different user classes.

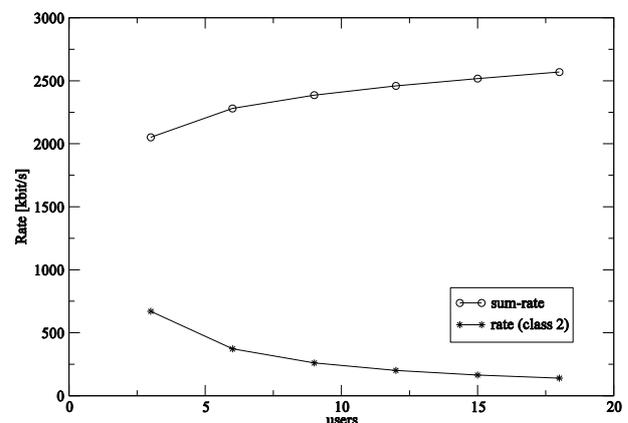

**Fig. 11**. Sum rate and per-user rate (user in class 2) vs. number of users, for a system with users at equal distance from BS.